\documentclass[aps,prd,groupedaddress,psfig]{revtex4}
\usepackage{graphicx}

\begin{document}



\title{At what particle energy do extragalactic cosmic rays start to predominate?}
\author{Tadeusz Wibig}
\affiliation{Experimental Physics Dept., University of \L \'{o}d\'{z}
}
\affiliation{ The Andrzej So\l tan Institute For Nuclear Studies,
Cosmic Ray Lab., \L \'{o}d\'{z}, Uniwersytecka 5, POB 447, \L \'{o}d\'{z} 1; Poland}
\author{Arnold W. Wolfendale}
\affiliation{Physics Department, University of Durham, Durham , UK.}

\begin{abstract}
We have argued (e.g. \cite{1}) that the well-known `ankle' in the
cosmic ray energy spectrum, at logE (eV) $\sim$ 18.7-19.0, marks
the transition from mainly Galactic sources at lower energies to
mainly extragalactic above.  Recently, however, there have been
claims for lower transitional energies, specifically from logE
(eV) $\sim$ 17.0 \cite{2} via $17.2-17.8$ \cite{3} to 18.0
\cite{4}.  In our model the ankle arises naturally from the sum of
simple power law-spectra with slopes differing by $\Delta \gamma
\sim 1.8$; from differential slope $\gamma = -3.8$ for Galactic
particles (near logE = 19) to $\gamma \sim -2.0$ for extragalactic
sources. In the other models, on the other hand, the ankle is
intrinsic to the extragalactic component alone, and arises from
the shape of the rate of energy loss versus energy for the
(assumed)
protons interacting with the cosmic microwave background (CMB).\\
Our detailed analysis of the world's data on the ultra-high energy
spectrum shows that taken together, or separately, the resulting
mean sharpness of the ankle (second difference of the
log(intensity$\times$E$^3$) with respect to logE) is consistent
with our `mixed' model.  For explanation in terms of extragalactic
particles alone, however, the ankle will be at the wrong energy --
for reasonable production models and of insufficient magnitude if,
as seems likely, there is still a significant fraction of heavy
nuclei at the ankle energy.

\end{abstract}
\maketitle

\section{Introduction}

A key property of the cosmic radiation is the division between those particles of Galactic
origin and those which are extragalactic.  It is virtually certain that those below about
logE (eV) = 17 are Galactic (G) and that the very highest energy particles are extragalactic
(EG), but the energy (E$_{0.5}$) at which EG starts to dominate is debatable.  It is this question
that is the subject of this paper.

In every model the Galactic magnetic field progressively loses its trapping power as the
rigidity increases so that the mean charge of the detected Galactic particles increases.
Ideally, measurements of the mass composition should be possible and these would appear to
indicate E$_{0.5}$ but there is the distinct likelihood of a significant fraction of heavy nuclei
in the EG beam.

Here, we start by considering the measured spectral shape and then examine the possibility
of an explanation in terms of primary protons.  This is followed by an analysis of the mass
composition and its relevance to the Galactic/Extragalactic boundary.

\section{The measured spectral shape}

A number of measurements have been made over the years, using increasingly large extensive
air shower (EAS) arrays.  Our earlier work \cite{1} gave a summary and here we update the analysis.
Figure 1 gives the individual spectra.  It will be noted that, when the energy range covered
is large enough, there is clear evidence for the presence of an ankle.  Defining `sharpness'
as in related work, (e.g. \cite{5}), by S = $\partial^2(\log IE^3)/\partial(logE)^2$ with a bin width of
logE = 0.2 -- or
converted to that value -- the values derived by us are as given in Table 1 (references to
the individual arrays are given in Ref.~\cite{1}).

A variety of methods have been used to determine the sharpness values including a variation
where the regions near the individual ankles ($\pm$0.2 in logE) are not used in determining the
actual ankle position in each case.  The result is that there should be no artificial sharpening
of the overall ankle (a situation suggested by some, e.g. \cite{6}).

Figure 2 shows the superimposed spectra normalized to the datum
intensity \cite{7} and the ankle-energy.  It is evident that the
well known large systematic differences in intensity do not occur
until beyond the ankle region.  Figure 3 shows the combined data
-- and corresponding uncertainties.  The extent to which our
simple explanation in terms of the addition of a rapidly falling
G-spectrum and a standard (approximately E$^{-2}$) EG-spectrum
fits the data near the ankle is clear.  The expectation (sum) has
observational uncertainties added to it; their effect is very
small.

Returning to the sharpness values, set (3) in Table 1 is probably
the safest to use. The mean value is S = $0.87 \pm 0.02$ where the
`error' has been derived from the dispersion of the individual
values.  The expected value for the `intersecting lines' from G
and EG spectra, is S = 0.96 without experimental resolution,
falling to $\approx$0.90 with the inclusion of noise. Thus, there
is consistency.

\section{Models for explaining the spectral shape in the ankle region}

As remarked earlier, the competing models have the Galactic flux reduced to a
very low level by the time the ankle is reached and the ankle is formed by the
effect of e$^+$e$^-$ production on the (assumed) EG proton spectrum.  Figure 4 shows
the Galactic fraction for the various models, including our own.  Figure 5 shows
the rate of energy loss for e$^+$e$^-$ and pion production from our own calculations
\cite{1} and by others \cite{8,9}.  If a function of this type is multiplied by an assumed
EG injection spectrum then the usual form of EG-proton spectrum is derived; it is
evident that there is, indeed, an ankle in the region of log E = 19. Figure 6
shows the multiplying factor for a number of situations, to be described.

It should be pointed out that the use of the factor taken simply as the reciprocal
of Figure 5, is for a Euclidian Universe, ie with no expansion and integration
for a uniform spatial and temporal distribution of sources out to the Hubble
radius.  In fact, there will be a number of factors which perturb the simple
form of the predicted spectrum, particularly at and below several times
log E = 19 (at higher energies the collection volume is local and expansion
effects are small).  The factors of concern are:
\begin{itemize}
\item[1]
The limit on red-shift beyond which the spatial density of galaxies -- which are
potential sources of ultra-high energy cosmic rays (UHECR) -- becomes small.
\item[2]
 Scattering of particles, which increases the path length and thereby travel time.
\item[3]
 Possible $z$-dependent production rates (the `cosmological increase' -- e.g. \cite{10})
\item[4]
 The enhanced CMB temperature, and consequent increased energy loss rates for the
protons, as $z$ increases.
\item[5]
The stochastic nature of the sources in space and in time.
\end{itemize}
Concerning 1, there is the immediate problem relating to the actual nature of the sources of UHECR.
In order of increasing general activity we have the following approximate densities, in units of
number per Gpc$^3$.

\begin{center}
\begin{tabular}{c|l}

Normal galaxies&
\ \ \ 3$\times 10^7$ out to $z \sim 2$\\ \hline
Colliding galaxies&
\ \ \ $3 \times 10^5$ at small $z$, increasing with $z$ \\ \hline
\ \ \ Seyfert galaxies\ \ \ &
\ \ \ $3 \times 10^5$ at small $z$, increasing slowly with $z$\\ \hline
Quasars&
\ \ \ Increasing with $z$, from $\sim$2 at $z = 0.5$ to $\sim300$ at $z \sim 2 $,
then falling \cite{11}
\end{tabular}
\end{center}

 For protons of
log E = 19 the attenuation length in a non-evolving universe is
$\sim 0.84 $Gpc (ie $z \sim 0.2$) and thus, only for quasars will
there be large stochastic effects.  The number of quasars in a
sphere of radius 0.84 Gpc will be $\sim$2 or 3 and at higher
energies the stochastic effects will even more serious, however,
here we are preoccupied with `ankle'-energies (ie $\approx$ log E
= 19).

If now we consider the increased energy losses as $z$ increases, the collection radius
will be smaller than 1.6 Gpc and the stochastic effects for quasars will be larger.
When allowance is made for the effect of large scale source density variations due to
superclusters (the `supercluster-enhancement', \cite{12}), voids and `great-attractors', the
spectral shape in the region of 10$^{19}$eV log E = 19 is unlikely to be smooth, if quasars
are responsible for the particles.  The chance of having the observed ankle and an
otherwise smooth spectrum is surely very unlikely.

Seyfert or colliding galaxies seem more likely here, although the temporal variations
for Seyfert galaxies (the fact that they are `on' for only about 1\% of the time) will
be significant.

Turning to factor 2, scattering is no doubt important for 10$^{19}$ eV protons.  The Larmor
radius in our usually adopted magnetic field of $3 \times 10^{-3} \mu$G \cite{13} is 3 Mpc and the
corresponding `scattering-horizon' is about 300 Mpc.  This will increase the stochastic
fluctuations somewhat, but this effect is unlikely to be large.

As remarked already, Figure 6 shows the factor by which the spectral shape should
be multiplied in order to achieve the expected shape.  The `predictions' are given
for a variety of situations, and can be considered in turn:
\begin{itemize}
\item[a] Uniform production model.  This is the standard, very approximate, case where
it is assumed that there is uniform production everywhere but $z$-dependent losses are
ignored. It is taken from the inversion of PW in Figure 5; that derived from the
other curves in the Figure would not be very different.  It will be noted that 
there is, indeed, an ankle near log E = 19 in the Euclidian case.

\item[b] Model with uniform production but allowing for
$z$-dependent losses.  We give the results \cite{8,9} for the
expected energy spectra for sources out to different maximum
$z$-values, with no source evolution.

\item[c]
An estimate for an injection spectrum which has the form E$^{-2}$ but has an
intensity dependence similar to that for radio sources, ie a `cosmological
increase' at `large' $z$ \cite{10}.  The effect of the increase is to raise the
intensity of lower energy particles which do not suffer much loss on the CMB.
\end{itemize}

\section{Discussion of the spectral shapes}

The first matter to discuss is the actual energy at which the ankle occurs.
This depends on the absolute calibration of the energies for which, as has
been remarked, there is no agreement.  Figure 2 shows the ankle at
log E = 18.93 for the Hi-Res \cite{7} normalization; with the AGASA \cite{1}
normalization the knee moves up to log E = 19.06.  Here we will take
log E = 19.0, the mean.

Turning to Figure 6 there is seen to be a wide disparity of shapes but
all have an ankle.  It is evident that BG$_{0.33}$ and PW have ankles at somewhat
too high an energy but, insofar as galaxy-sources of UHECR almost certainly
extend further than $z = 1$ this cannot be regarded as fatal for the EG-model.
For PW, the maximum sharpness is: \\
S$_{PW} = 1.1$, i.e. not inconsistent with observation, although, as
remarked, the energy value is high and the physics is suspect.

For the more likely case of a $z$-dependent injection rate (eg `PJ')
the sharpness is reasonable (S$_{PJ} = 0.75$) but the energy at which the
maximum occurs, log E$\approx$18.6, is too low.

Before continuing to examine the relevance of the mass spectrum
other considerations are necessary, related to the other factors
referred to in section 2.  The smoothing introduced by factor 5 in
section 2, viz stochastic processes is important.  One such
relates to the non-uniformity of density of galaxies in the
universe.  Calculations allowing for the enhanced density of local
galaxies in the VIRGO supercluster (`supercluster enhancement'
\cite{12}) show a displacement of the patterns in figure 6 to
higher energies by a factor $\Delta$log E$\approx$0.2, thereby
taking `PW' even further from `the truth'.

An examination of the topology of galaxies as a whole shows the presence
of remarkable fluctuations in density on scales even higher than that of
superclusters.  For example, there is evidence for `huge regions of
underdensity in both hemispheres' \cite{15} and there is a 30\% reduction
to $z = 0.1$ ($\approx$440 Mpc) in the Southern Hemisphere.  The net result for
the range of distances corresponding to the ankle ($\approx$840 Mpc for log E = 19)
is that we estimate a standard deviation in log E of about 0.3.
Application of a Gaussian with this standard deviation causes a reduction
in sharpness by a factor of $\sim$0.6.  The value for PW then become lower
than needed.

The conclusion is, therefore, at this stage, that the position of
the ankle allows an explanation in terms of EG particles alone,
but that its magnitude is rather low (0.6 compared with a needed
0.9).  Furthermore, a cosmological increase in Galactic output --
which seems likely on general astrophysical grounds -- is not
allowable, nor indeed is allowance for $z$-dependent losses, which
must be present. The decisive blow against the EG origin of the
ankle comes from the mass composition, however, and this aspect
will now be examined.

\section{The mass composition of the particles above logE (eV) = 17}

The relevance of the mass composition here is that we have assumed, so far,
that at least at energies above log E = 19, the bulk of the EG particles are
protons.  It has proved notoriously difficult to determine the mass spectrum,
largely because of uncertainties in the high energy interaction model to be
adopted so far above the (accelerator) energies at which the models can be
checked, but some progress has been made.

It would be expected that Galactic particles would be increasingly lost as
the particle rigidity increases and, in consequence, that the fraction of
heavy nuclei (iron) would continuously increase.  An indication of the extent
of the Galactic component is therefore tied up with the fraction of the beam
which we assume to be composed of such heavy nuclei.

Over the years we have studied this aspect in a number of ways, principally
by examining the frequency distribution of depth of shower maximum and by
studying the data relating to the `Galactic Plane Excess'.

Figure 7 shows the results for $f_{Fe}$, the fraction or iron in the primary beam.
The results denoted `$f$' are from the frequency distribution of $X_{\rm max}$  values as
analysed by ourselves \cite{16}; we prefer this technique to that of using the mean
(e.g. \cite{7}) because of the effect of uncertainties in the nuclear interaction model.
In the fluctuation model, the high mass `tail' to the distribution is assumed
due to iron and the distribution displaced accordingly.  The results from the
mean are considered later.

The results marked `T' come from our analyses of the b-distribution in the
Inner Galaxy using trajectory calculations for particles accelerated in the
Galactic Plane \cite{17}.

We turn now to other estimates, from the mean depth of shower maximum.  A
useful survey \cite{18} gives the median denoted D in Figure 7; this includes
the low values from Hi-Res. (II) which provided the reason for the low
cross-over point given earlier.  (Figure 4).  Also included is the latest
AGASA measurement \cite{19}, at log E = 19.3, denoted `A'.

It is evident that the new determinations (D and A) are, if anything,
somewhat higher than our own, particularly above log E = 18.5, so that
our estimate of the Galactic fraction in Figure 4 has a measure of
confirmation.  The values at log E = 19 can be taken as an example.
The overall mean of $f_{Fe}$ (Figure 7) is at 0.35 so that, for two components
only, Galactic iron and Extragalactic protons, the corresponding Galactic
fraction is also 0.35.  This can be compared with our `need' of 0.35 for
the Hi-RES normalization and 0.42 for AGASA.  There is clearly no problem
in terms of consistency.  However, there is an important implication for
the expected sharpness, as will now be described.

\subsection{Sharpness for a mixed composition}

Insofar as the energy-loss rates for protons -- and nuclei (eg iron) are different,
their sum will have a different shape to either.  Figure 8 shows the equivalent to
Figure 7.  For the best estimate of the p/Fe mixture of 90/10, the sharpness is
very small, and at too low an energy.

\section{Conclusions}

A variety of evidence leads us to believe that the ankle in the
primary energy spectrum is due to the rapid transition from
Galactic to Extragalactic particles and is not due to
Extragalactic particles alone, which are required to predominate
above logE = 18.0.  In our model, at logE=19.0, 30\%-- 40\% of the
particles are still Galactic, although this fraction is falling
rapidly with increasing energy.  In `competing models' this
fraction is 10\% for the Hi-Res case \cite{2}, 2\% in the
`Hillas-model' \cite{4} and zero in the Berezinsky et al. model
\cite{3}.

\acknowledgements{
  The authors are grateful to A.M.Hillas for useful comments,and for
making available the details of his model.}

\begin{figure}[th]
\centerline{
\includegraphics[width=6.1cm]{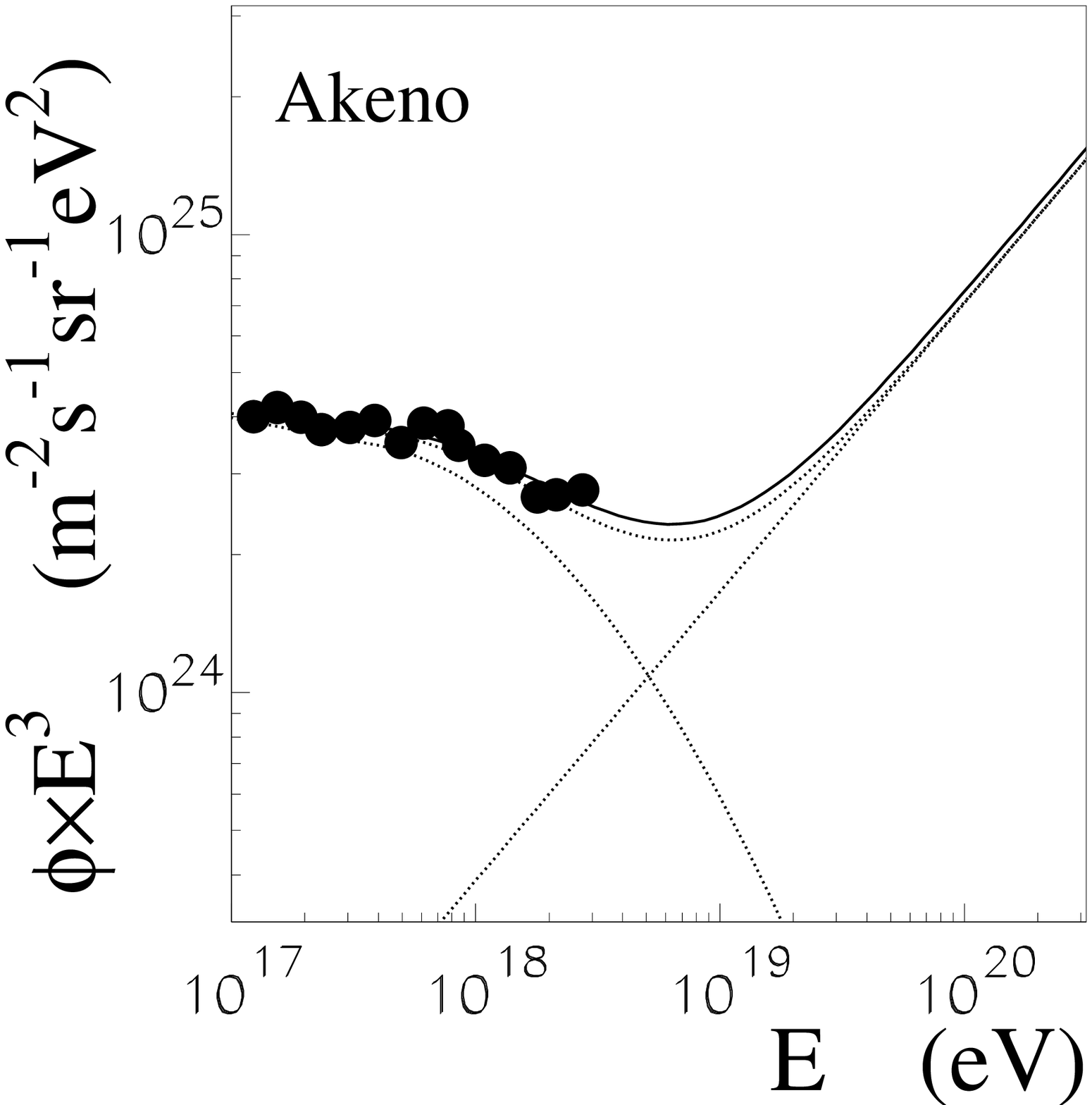}
\includegraphics[width=6.1cm]{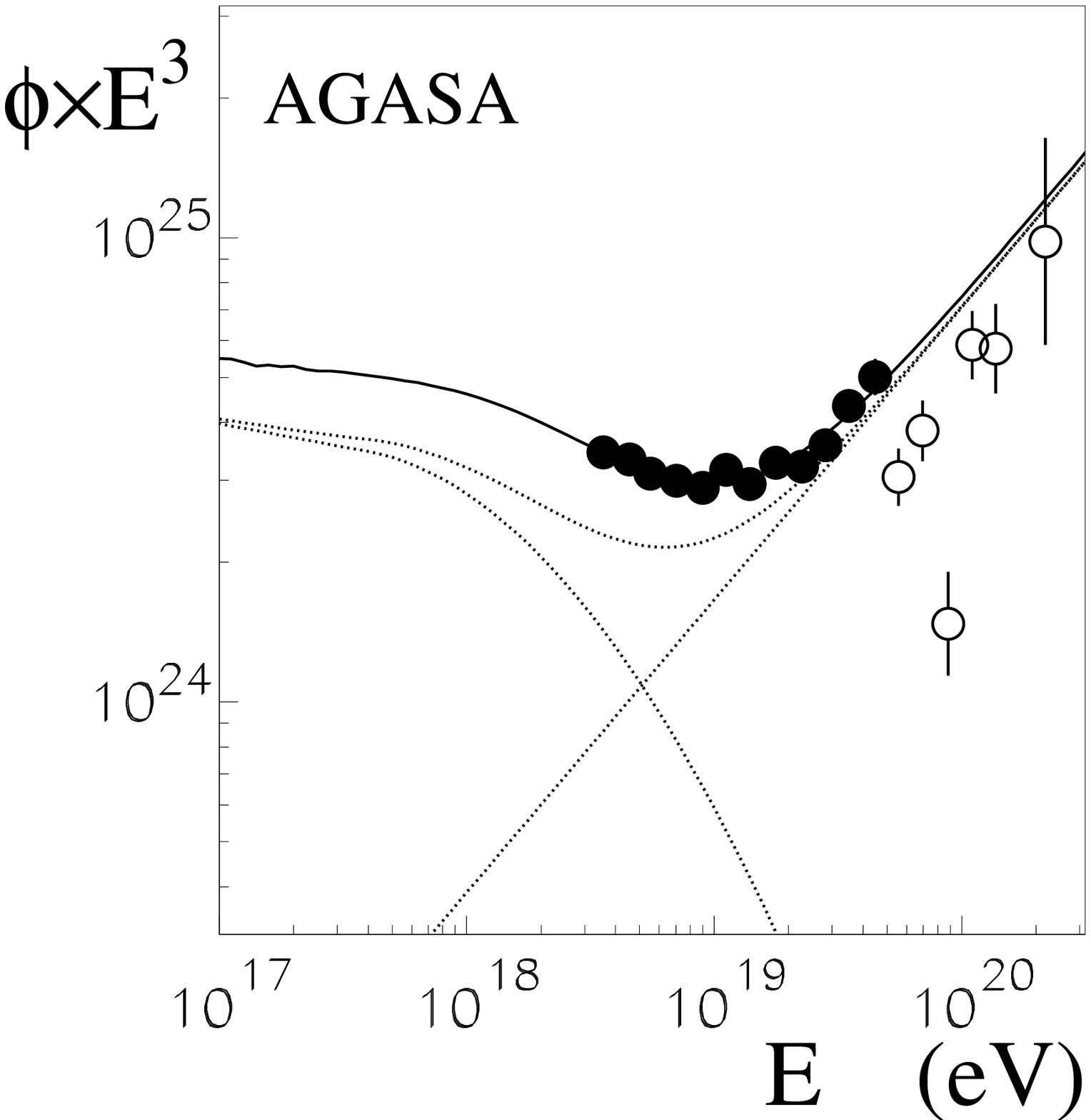}
\includegraphics[width=6.1cm]{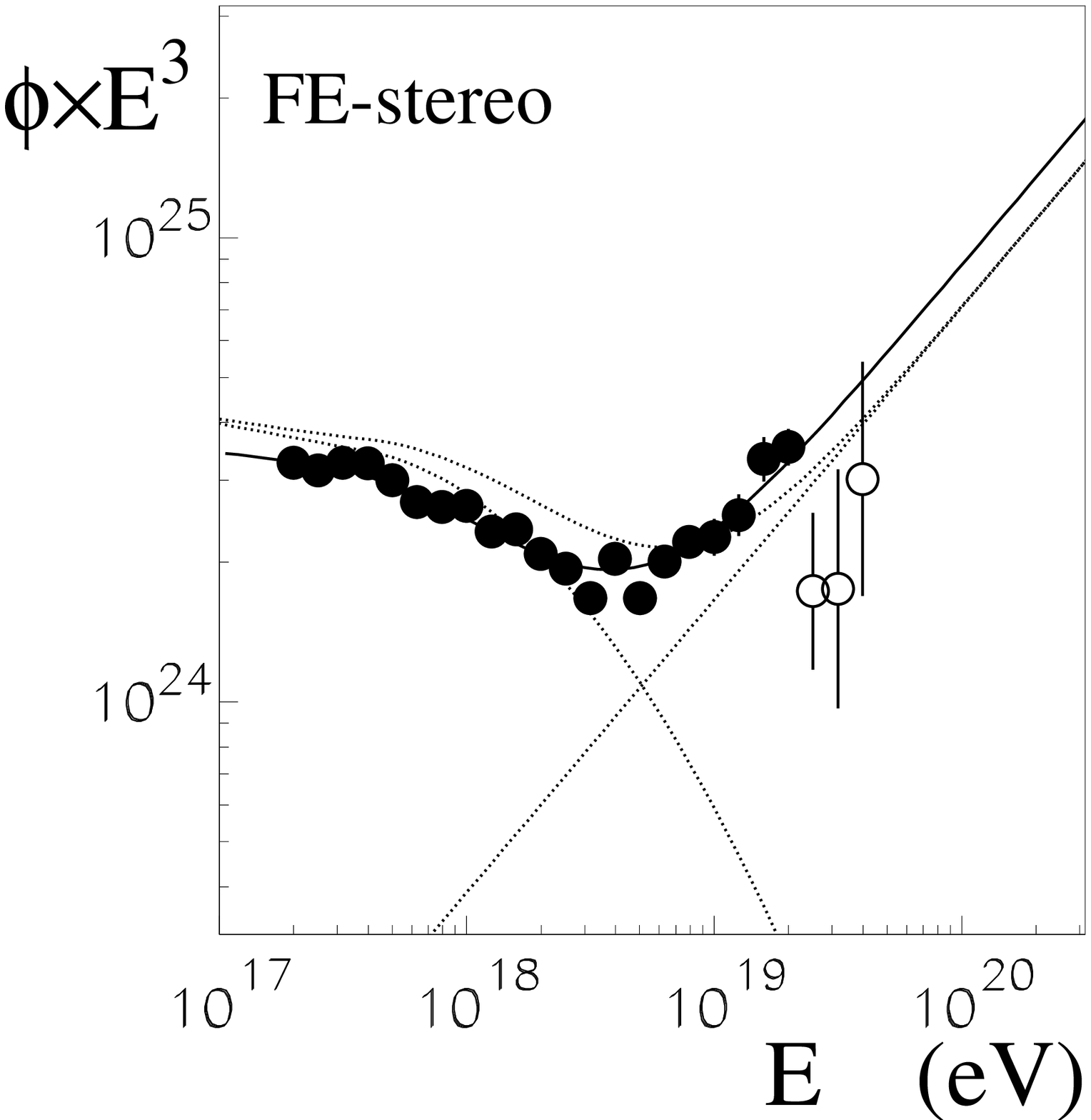}
}
\centerline{
\includegraphics[width=6.1cm]{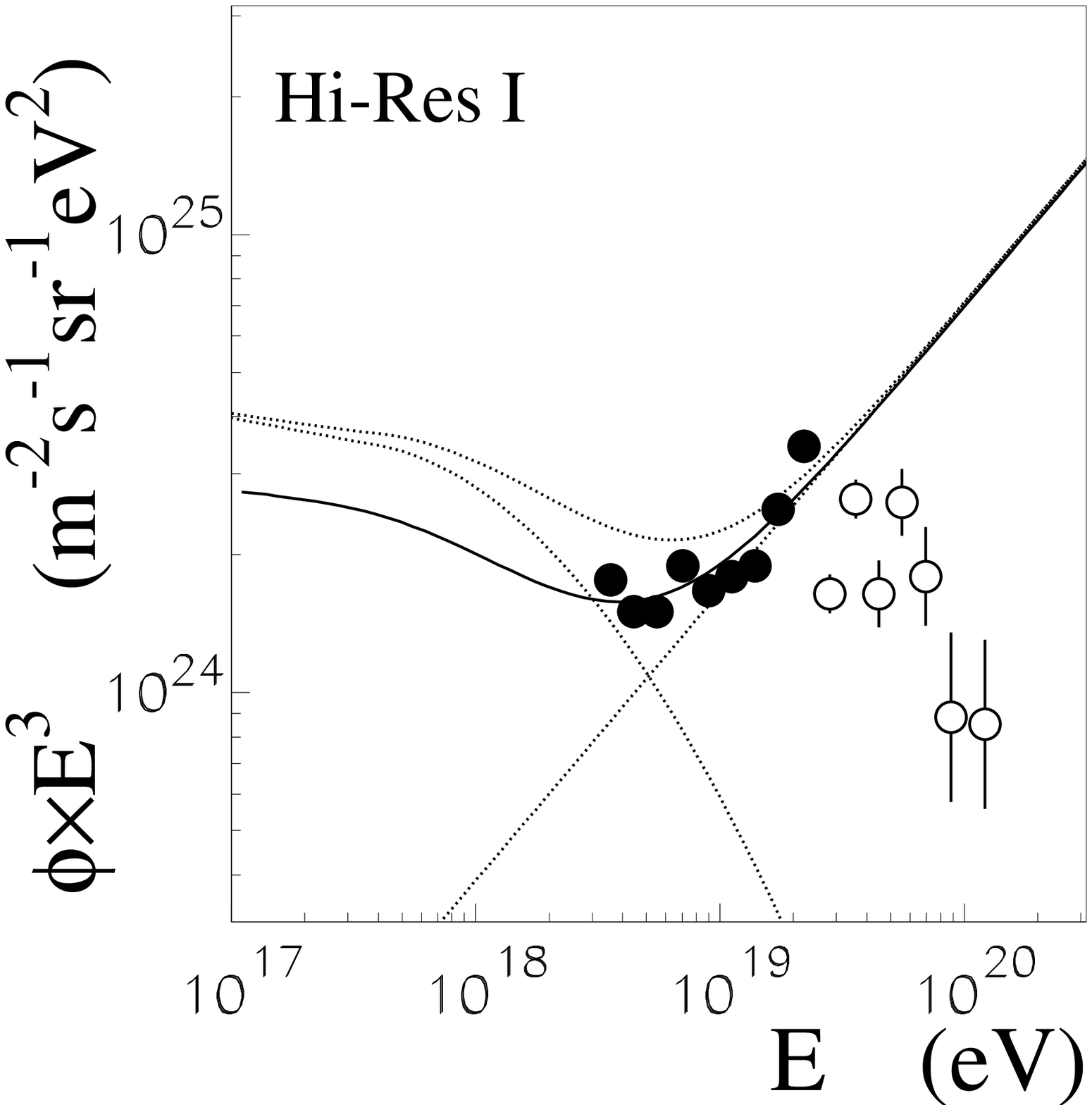}
\includegraphics[width=6.1cm]{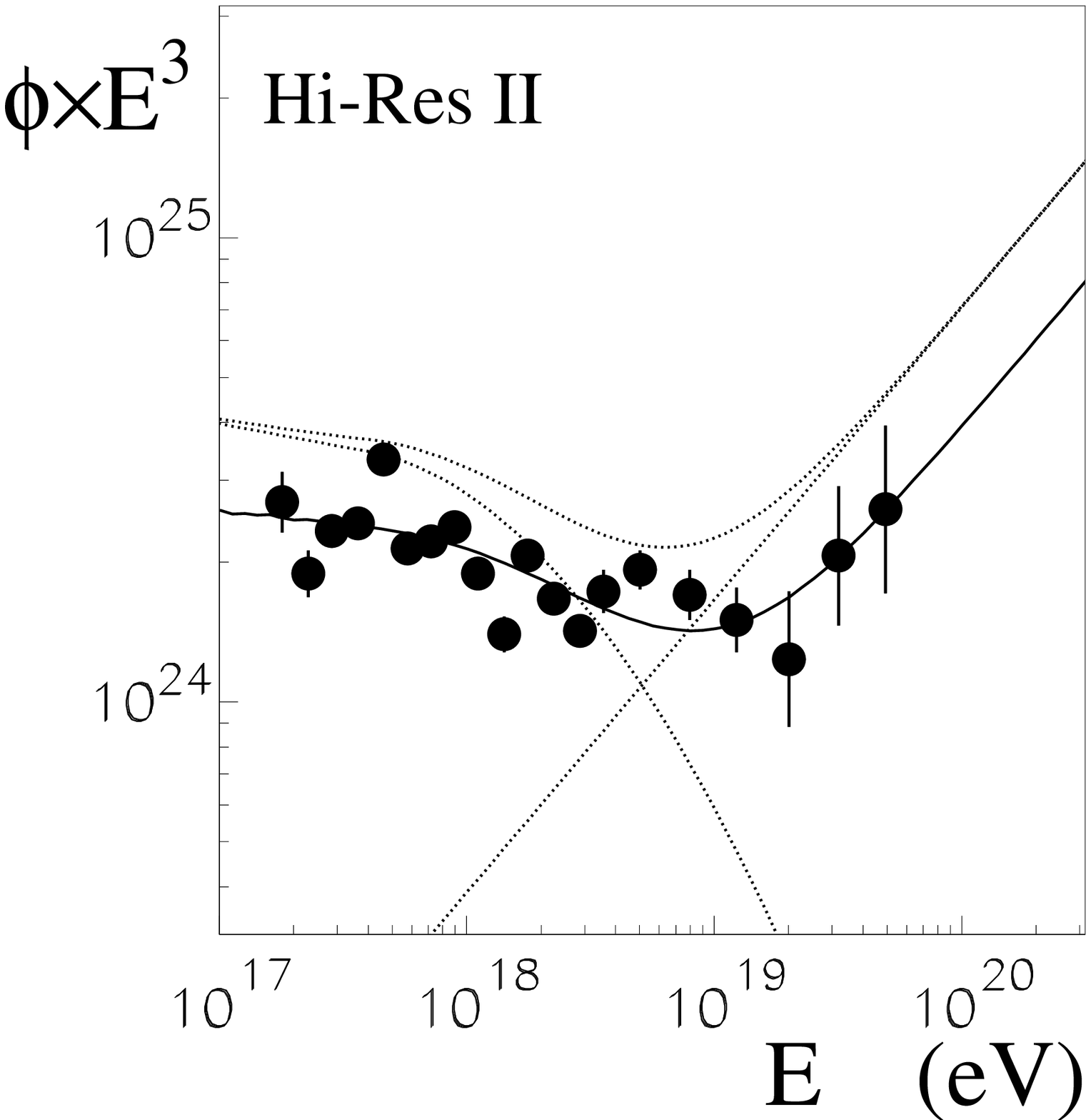}
\includegraphics[width=6.1cm]{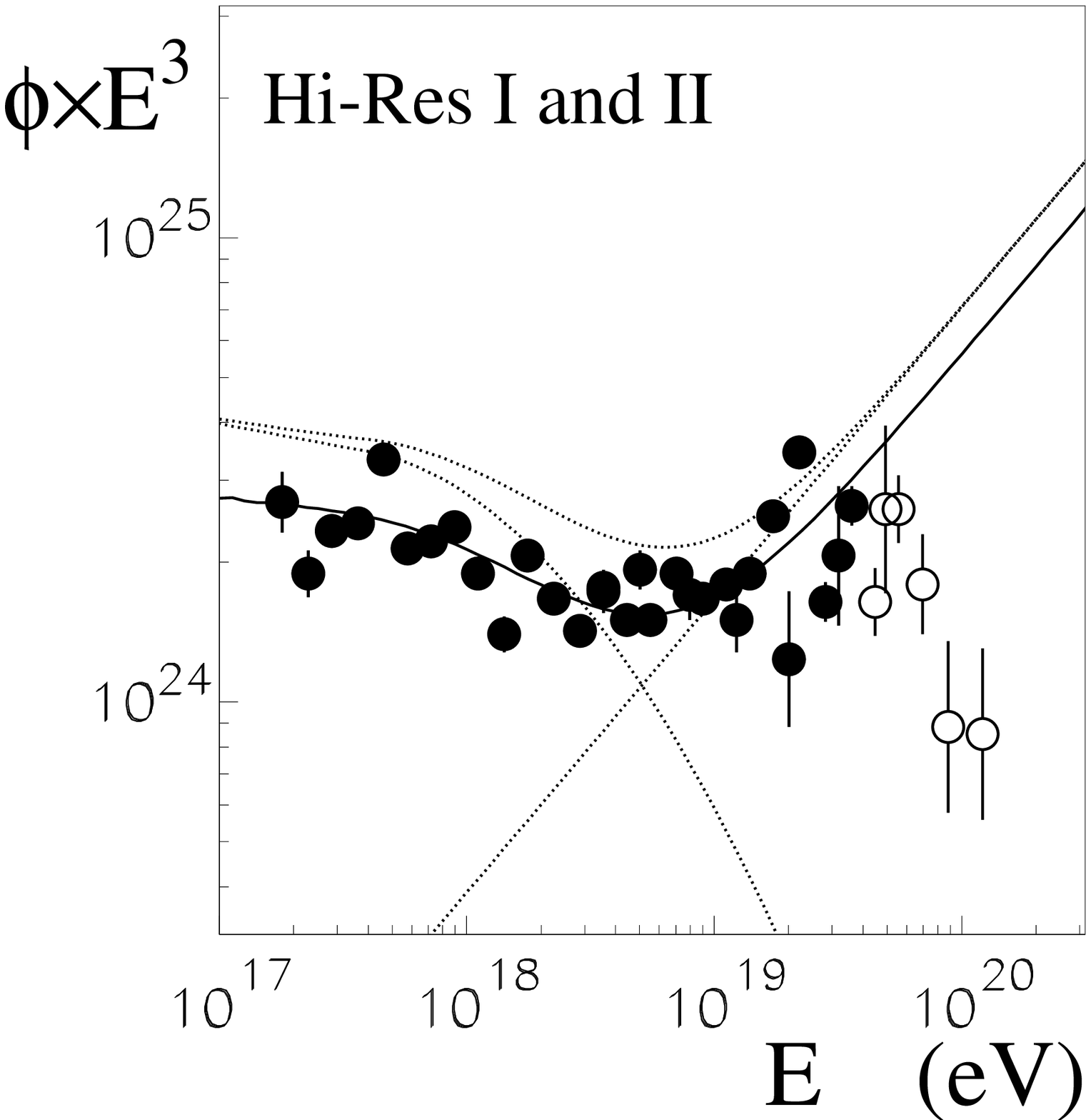}
}
\centerline{
\includegraphics[width=6.1cm]{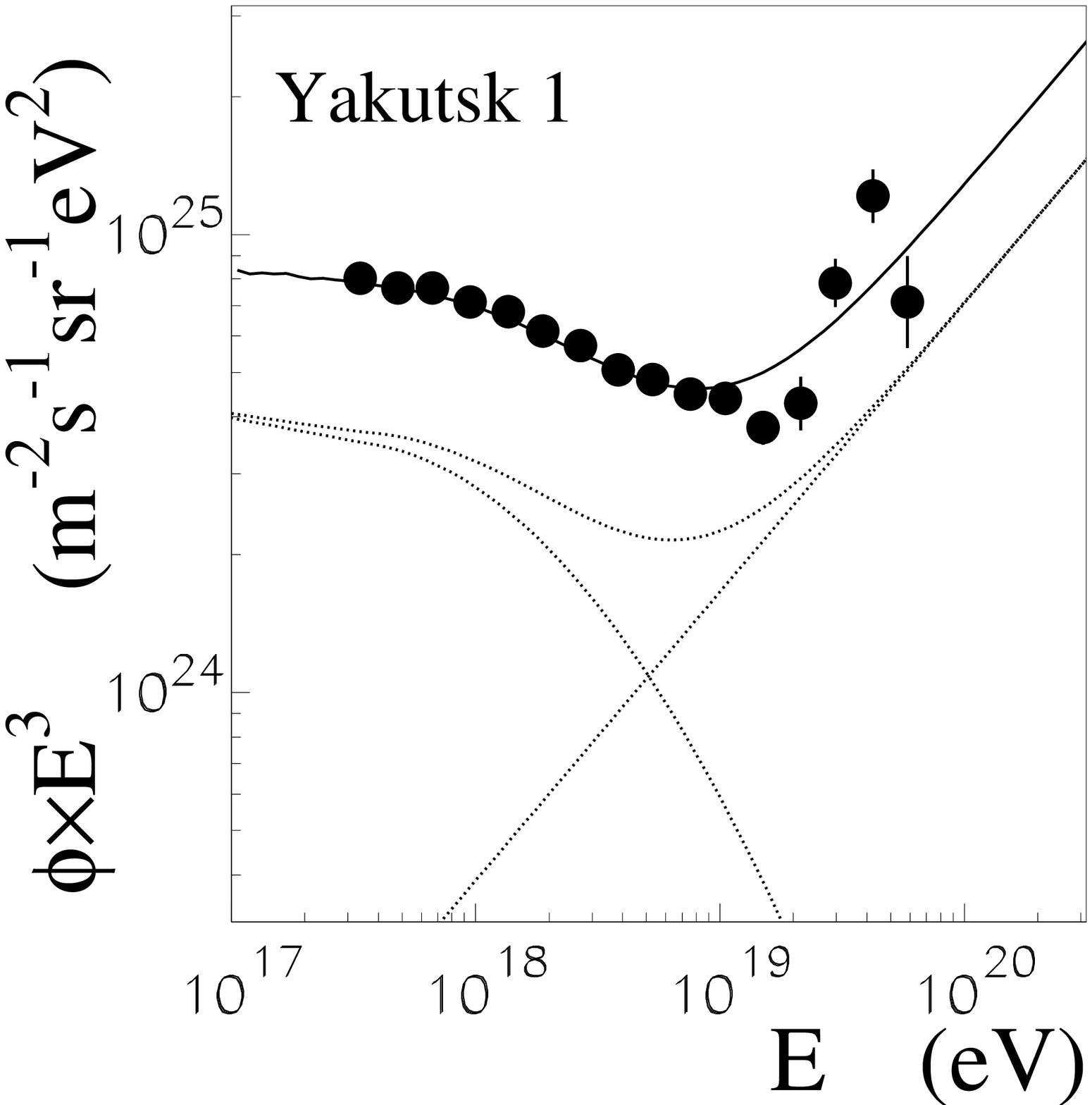}
\includegraphics[width=6.1cm]{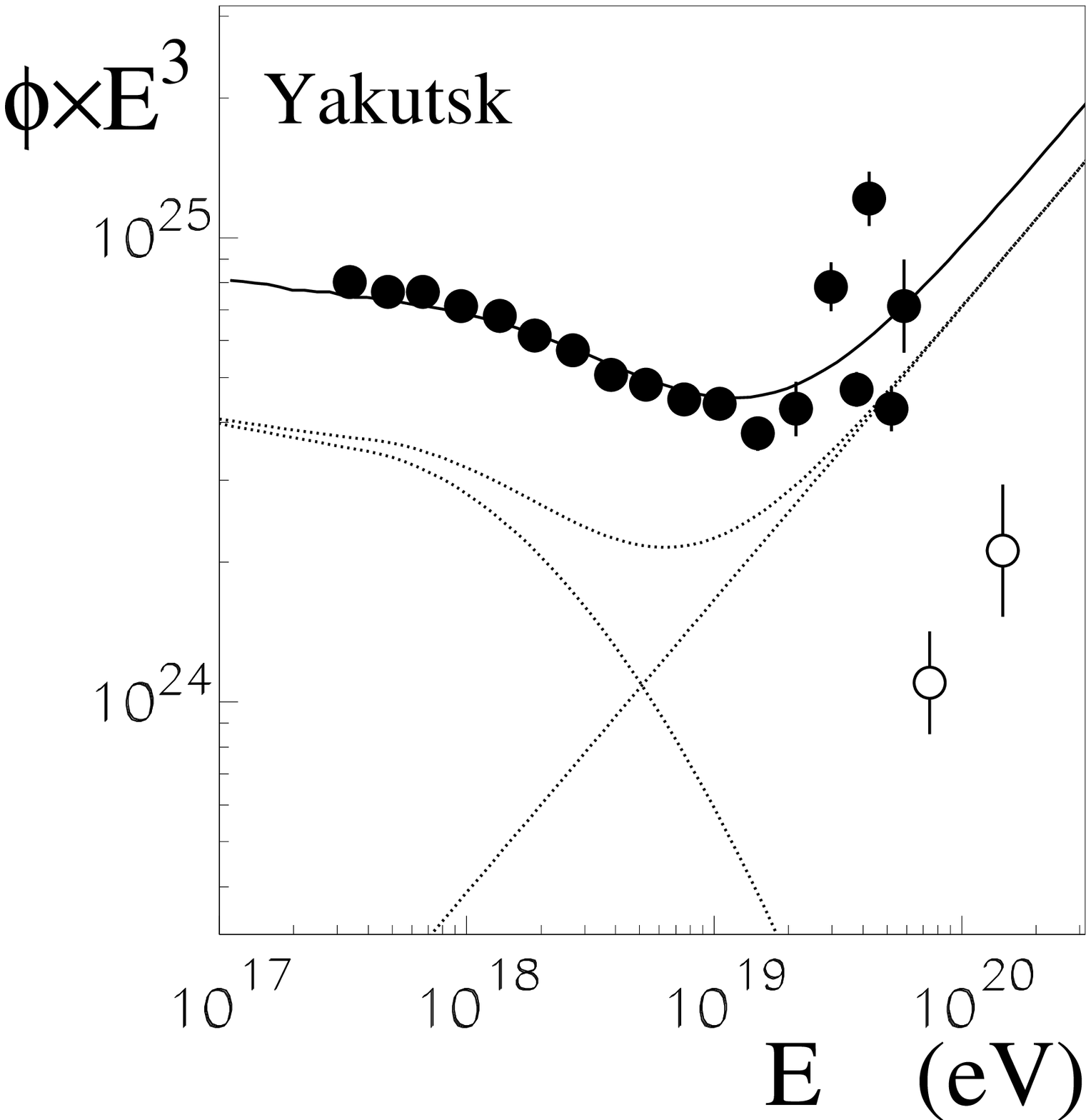}
\includegraphics[width=6.1cm]{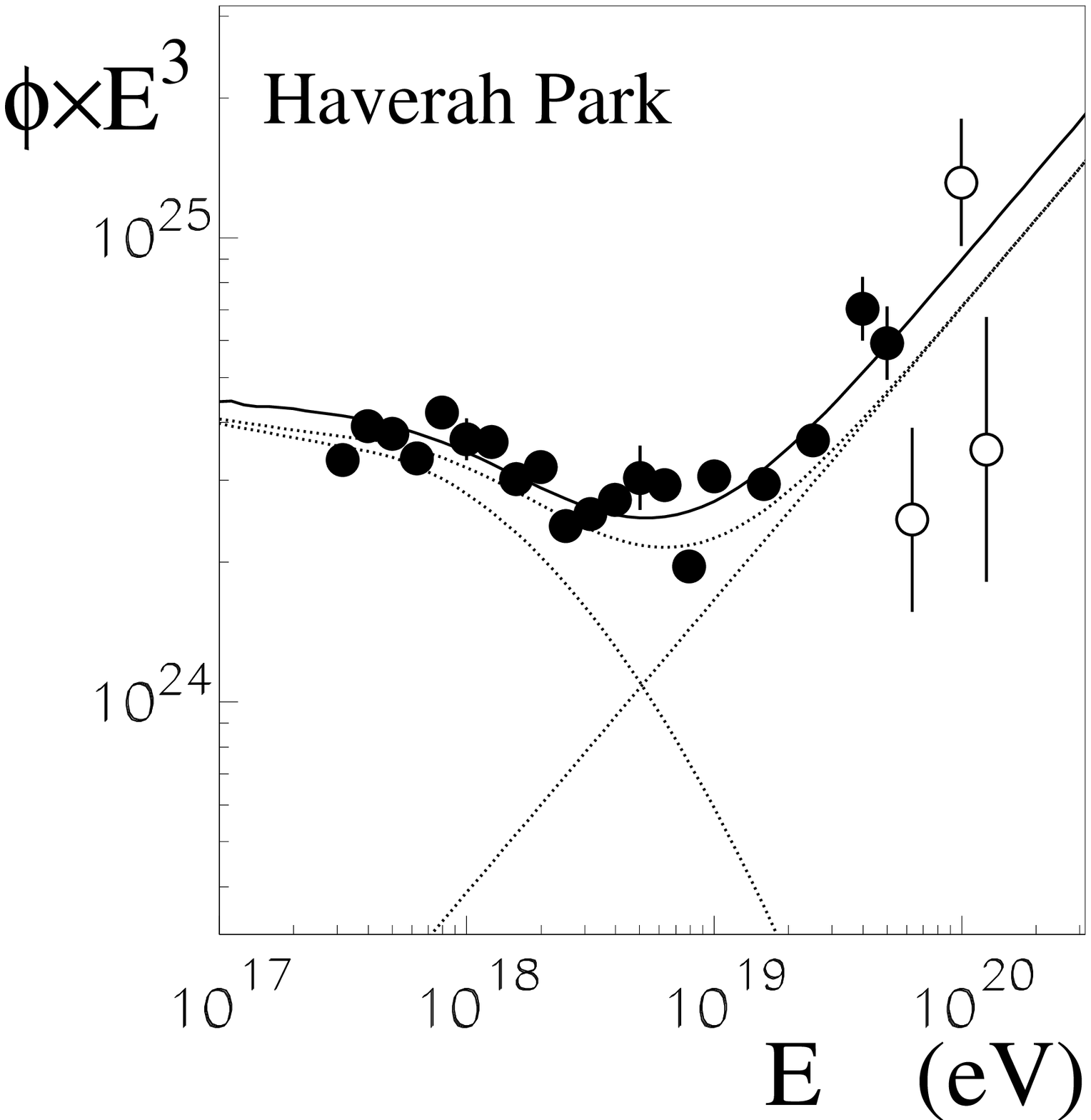}
 }
\vspace*{8pt}
\caption{Individual spectra for the major EAS arrays (for sources of data, see [1])
Thin dashed lines shows a smoothly falling Galactic spectrum and an EG
spectrum with differential slope $-2.37$ and their sum.
The solid line is the UHECR spectrum corrected for each individual array
resolution and systematic shifts.}
\end{figure}

\begin{figure}[th]
\centerline{\includegraphics[width=10cm]{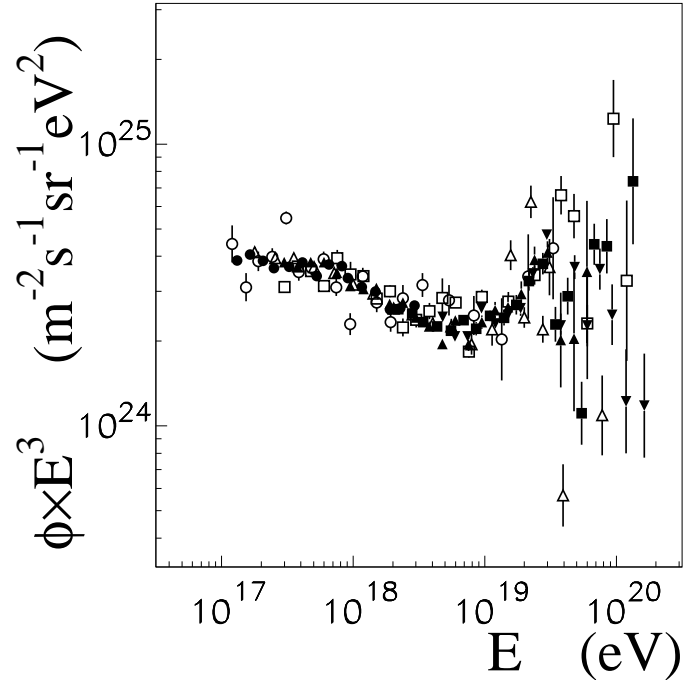}}
\vspace*{8pt}
\caption{The individual spectra normalized to a fixed intensity at the ankle.  It
is evident that the well-known considerable dispersion does not start
until above the ankle.}
\end{figure}

\begin{figure}[th]
\centerline{
\includegraphics[width=6cm]{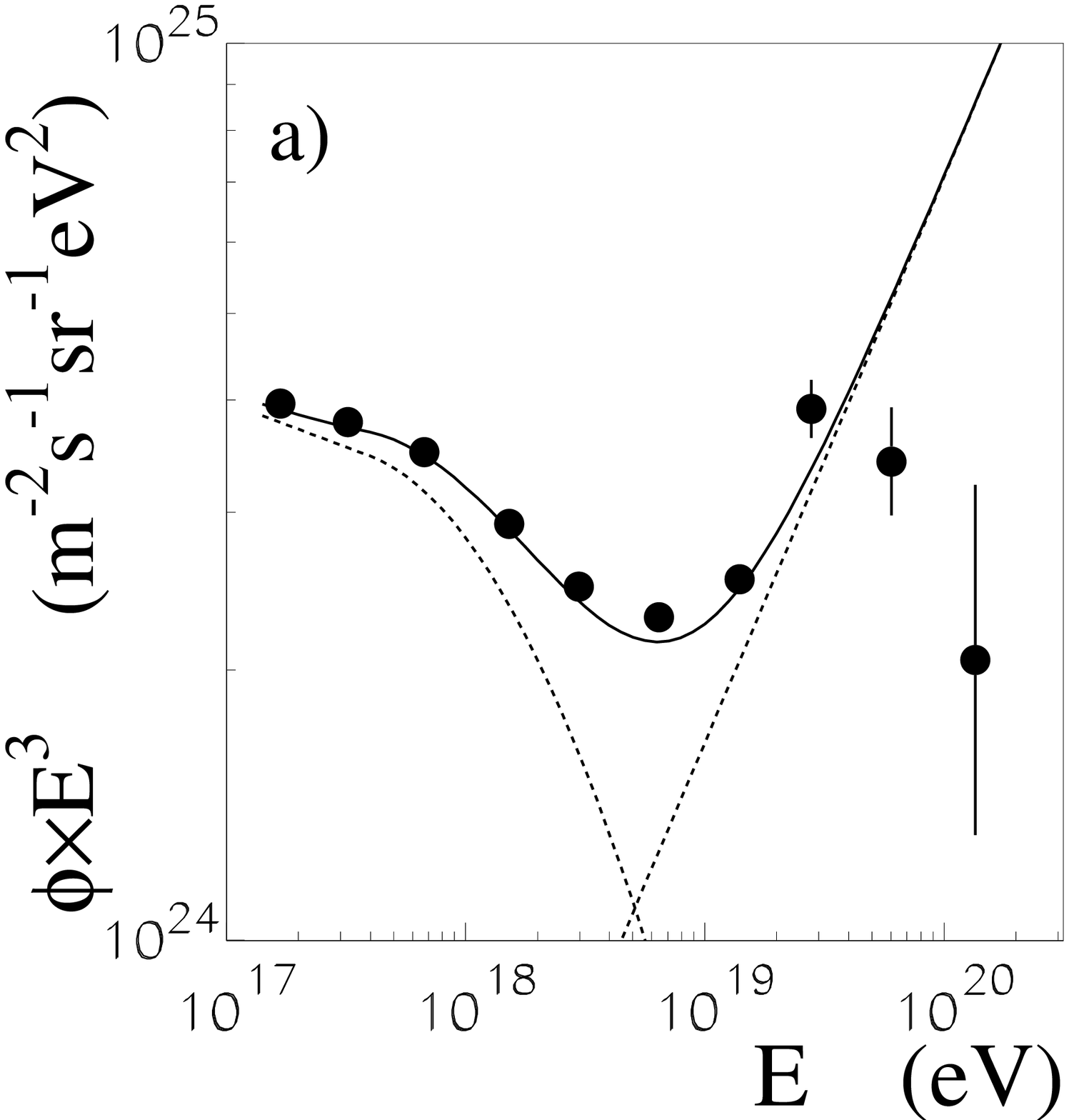}
\includegraphics[width=6cm]{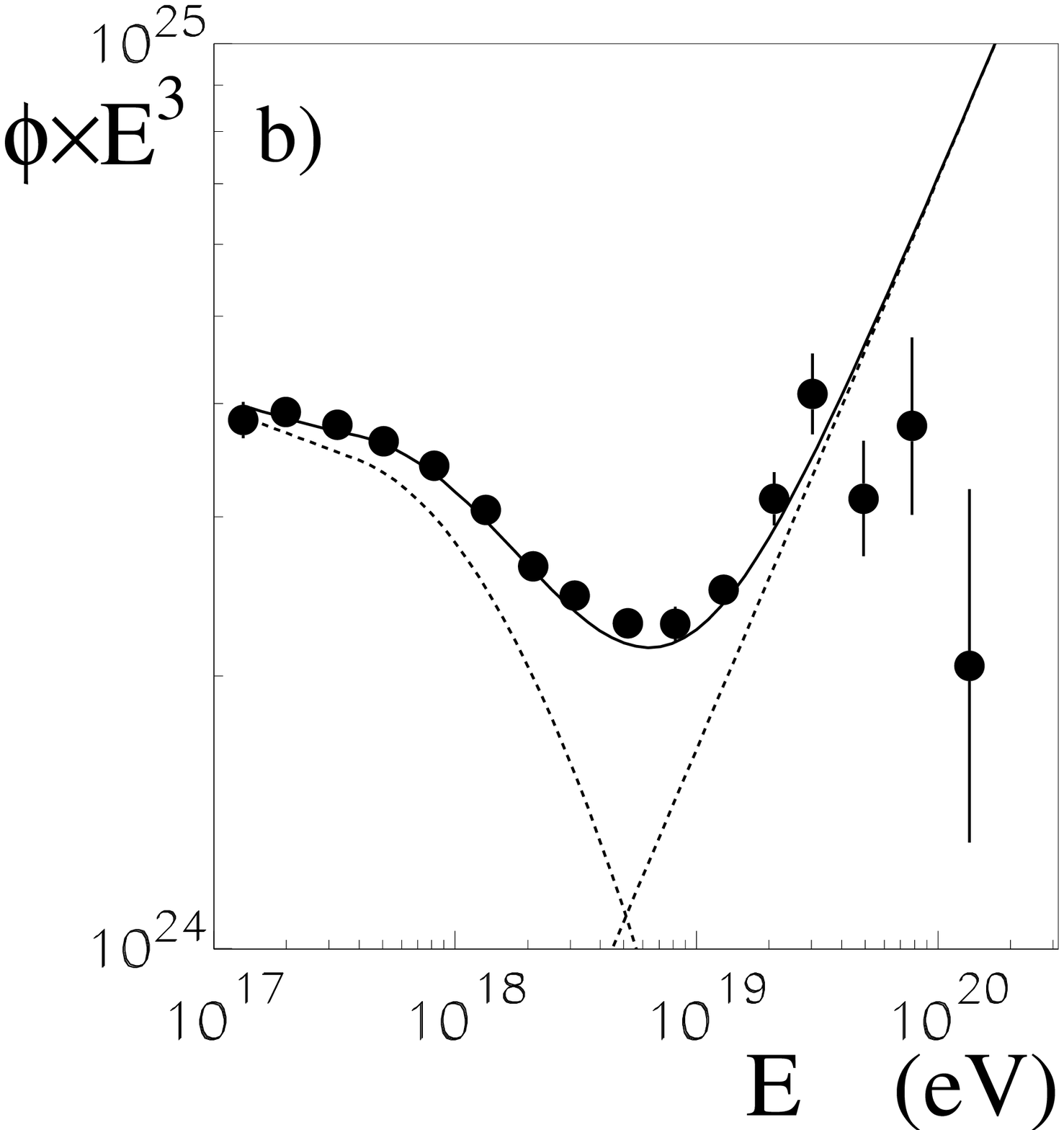}
}
\centerline{
\includegraphics[width=6cm]{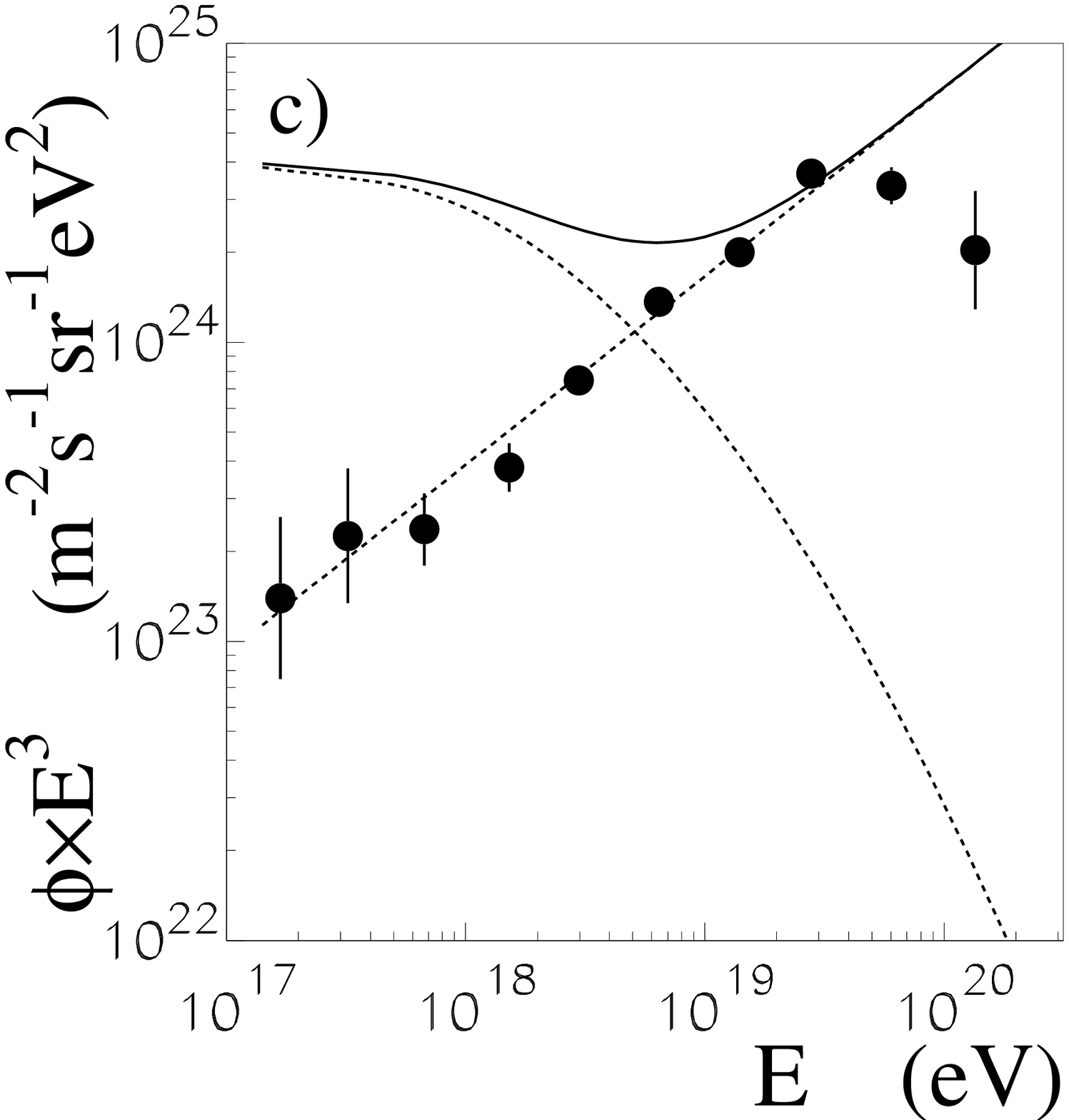}
\includegraphics[width=6cm]{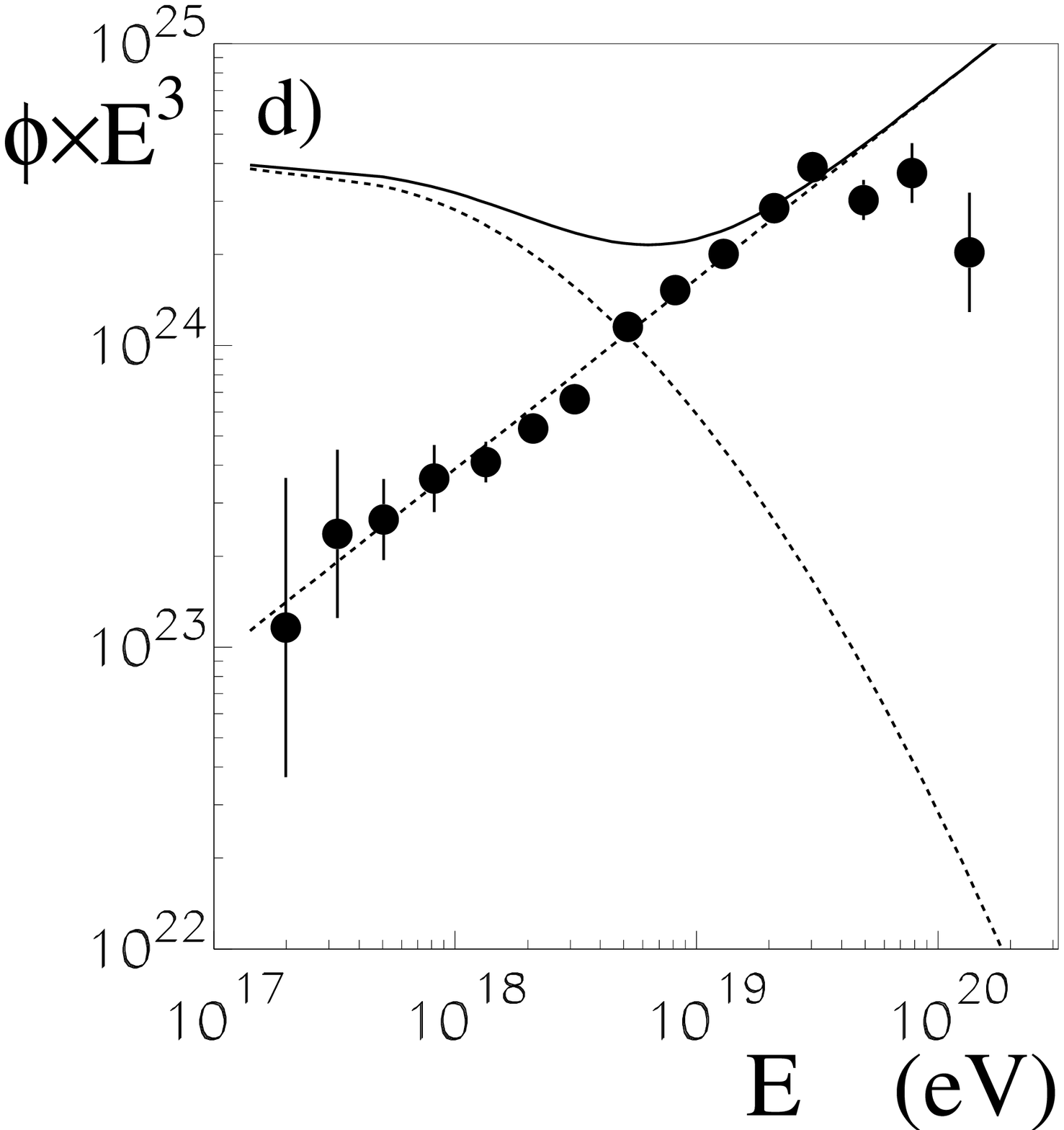}
}
\vspace*{8pt}
\caption{(a, b) A world summary of the UHECR with different binning 
(left and right).
Galactic and EG spectra (the latter with slope $-2.37$) are indicated, as is their
sum.
\\
(c, d) The world average data after subtraction of the Galactic component.  The
line corresponds to a spectrum with slope $-2.37$. It is evident that a smoothly
falling Galactic spectrum results in an EG spectrum that is a simple power law
for energies below log E $\sim$ 19.4.}
\end{figure}

\begin{figure}[th]
\centerline{\includegraphics[width=10cm]{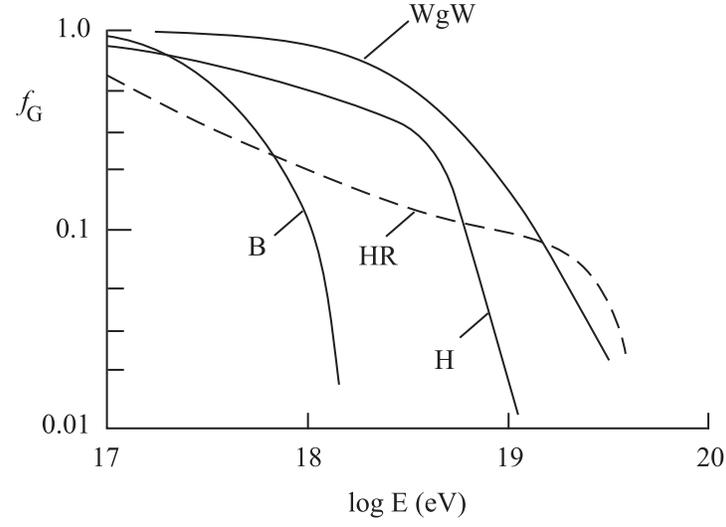}}
\vspace*{8pt}
\caption{Galactic fractions from the models of various authors:
        B: Bersinsky{\it et al.} [3],
        H: Hillas [4],
        HR: High-Res [2],
        WgW: Wibig and Wolfendale [1, 16].}
\end{figure}

\begin{figure}[th]
\centerline{\includegraphics[width=10cm]{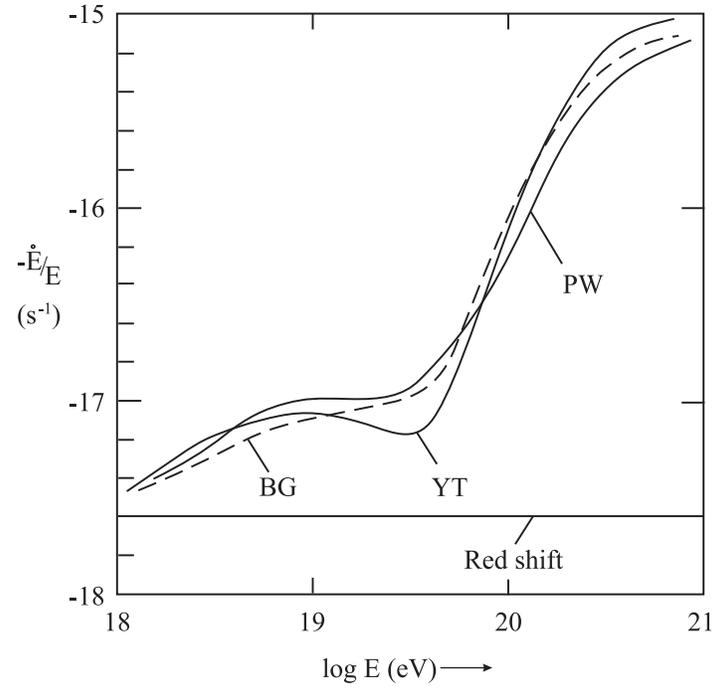}}
\vspace*{8pt}
\caption{Rate of energy loss of protons on the CMB from various authors:
PW: `present work',
BG: Ref. [8],
YT: Ref. [9].}
\end{figure}

\begin{figure}[th]
\centerline{\includegraphics[width=10cm]{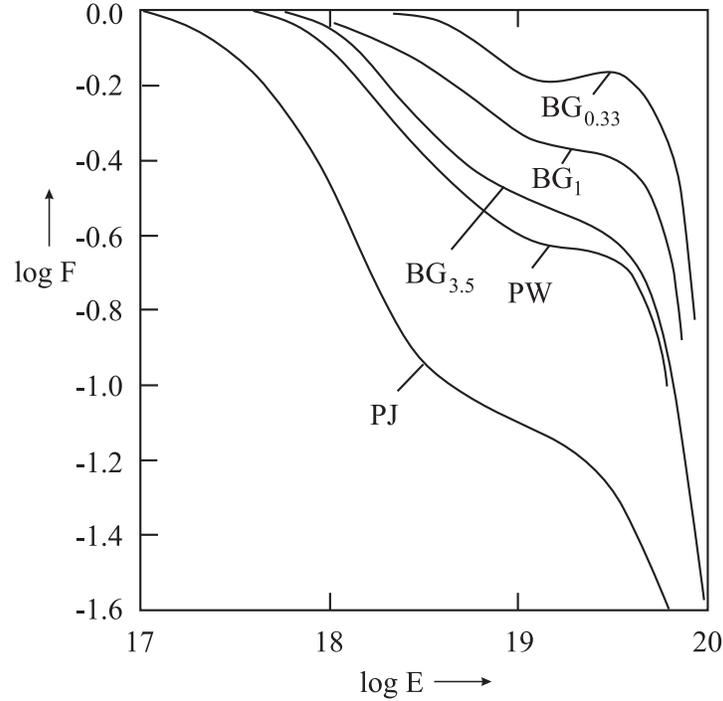}}
\vspace*{8pt}
\caption{Factor by which to multiply the injection spectrum for primary protons.\\
BG: Berezinsky and Grigor'eva [8] for no cosmological increase in injection
spectrum with increasing red shift, but integrating out to various maximum
$z$-values (eg BG$_{3.5}$ means integrating to $z_{\rm max} = 3.5$)
\\
PW: `present work' from Figure 5.
\\
PJ: Protheroe and Johnson [10]: this includes a cosmological increase in injection with increasing $z$.}
\end{figure}

\begin{figure}[th]
\centerline{\includegraphics[width=10cm]{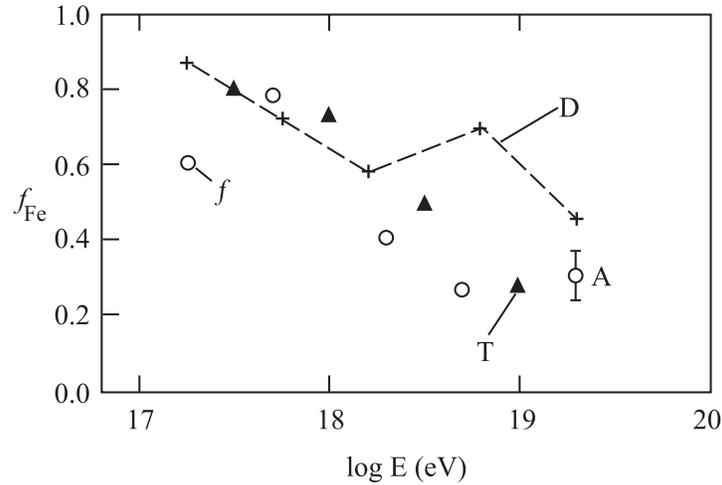}}
\vspace*{8pt}
\caption{Fraction of heavy nuclei (Fe) in the primary cosmic ray beam.
\\
Origin of the estimates:
       $f$: frequency of shower maximum values [16],
       T: `trajectories' for Galactic particles [17],
       A: Akeno [19],
       D: Dove{\it et al.} [18].
\\
There seems little doubt that there are heavy nuclei at log E = 19.}
\end{figure}

\begin{figure}[th]
\centerline{\includegraphics[width=10cm]{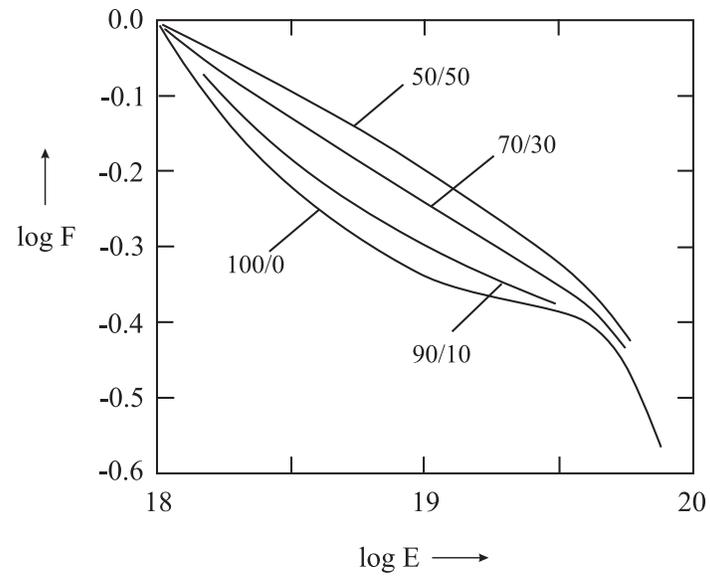}}
\vspace*{8pt}
\caption{Factor by which to multiply the injection spectrum
for a uniform distribution of sources in the universe
(cf Fig.6 PW) but with a mixture of particles, protons and iron,
in the ratios shown.  P/Fe = 70/30 is the best fit to the data
in Figure 7; it will be noted that there is no ankle here.}
\end{figure}

\begin{table}
\caption{Sharpness values for the major arrays.}
\begin{tabular}{l|c|c|c}
\multicolumn{1}{c|}{\bf array}&
\multicolumn{3}{c}{\bf sharpness}\\
\ \ \ \ \ \ \ \ \ \ \ \ \ \ \ \ \ \
&\ \ \ \ 1\ \ \ \ &\ \ \ \ 2\ \ \ \ &\ \ \ \ 3\ \ \ \ \\
\hline
Akeno$^*$&2.68&2.68&0.86\\
AGASA&0.99&1.03&0.85\\
Fly's Eye&1.06&0.98&0.93\\
Hi Res I+II&0.58&0.98&0.90\\
Haverah Park&0.62&0.83&0.91\\
Yakutsk 1+2&0.90&0.80&0.78\\
\hline
Mean S&0.83&0.92&\ 0.87$\pm$0.02 \ \\
\end{tabular}

\begin{tabular}{ll}
-------------------------- \\
\multicolumn{2}{l}{
$\ ^*$ \small
The Akeno values are uncertain and have not been used in the analysis.}
\end{tabular}

\vspace{.2cm}

\hspace{2.1cm}
\begin{tabular}{ll}
{\small key:}&\\
\ \ \ \small 1:& \small Fits to all the data points.\\
\ \ \ \small 2: &\small Fits excluding those within $\pm 0.2$ of the minimum.\\
\ \ \ \small 3: &\small As 2 but with one, universal index (EG exponent$ = -2.37$) for all experiments.\\
\end{tabular}
\end{table}

\end{document}